\documentclass[11pt,a4paper]{article}
\pdfoutput=1
\usepackage{jheppub}

\usepackage{mathrsfs}
\usepackage{booktabs}
\usepackage{slashed}
\usepackage{subcaption}
\usepackage{enumerate}

\usepackage{tikz}
\usetikzlibrary{decorations,trees}
\usetikzlibrary{decorations.markings}
\usetikzlibrary{patterns,calc}
\tikzstyle{every picture}+=[remember picture]
\tikzset{
	glu/.style={decorate,
		decoration={coil,amplitude=3pt, segment length=6pt}},
	fermion/.style={postaction={decorate},
		decoration={markings,mark=at position .55 with {\arrow{>}}}},
	scalarnoarrow/.style={dashed}
}
\newcommand{\lag}{\mathcal{L}}

\newcommand{\crmn}[2]{{C}_{#1}^{(#2)}}
\newcommand{\cscr}[2]{\mathscr{C}_{#1}^{(#2)}}

\newcommand{\mw}{M_\textsc{w}}
\newcommand{\axa}{A\otimes A}
\newcommand{\axv}{A\otimes V}

\graphicspath{{./figs/}}

\title{$\mathbf Z^\prime$-mediated Majorana dark matter: suppressed direct-detection rate and complementarity of LHC searches}

\author[a]{T. Alanne,}
\author[b]{F. Bishara,}
\author[a]{J. Fiaschi,}
\author[a]{O. Fischer,}
\author[a]{M. Gorbahn,}
\author[a,c]{and U. Moldanazarova}

\affiliation[a]{Department of Mathematical Sciences, University of Liverpool, Liverpool L69 3BX, United Kingdom}
\affiliation[b]{Deutsches Elektronen-Synchrotron DESY, Notkestr. 85, 22607 Hamburg, Germany}
\affiliation[c]{Faculty of Physics and Technology, Karaganda Buketov University, 100028 Karaganda, Kazakhstan}

\emailAdd{alanne@liverpool.ac.uk}
\emailAdd{fady.bishara@desy.de}
\emailAdd{fiaschi@liverpool.ac.uk}
\emailAdd{oliver.fischer@liverpool.ac.uk}
\emailAdd{mgorbahn@liverpool.ac.uk}
\emailAdd{psumolda@liverpool.ac.uk}

\abstract{
We study the direct-detection rate for axial-vectorial dark matter scattering off nuclei in an $\mathrm{SU}(2)\times \mathrm{U}(1)$ invariant effective theory and compare it against the LHC reach.
Current constraints from direct detection experiments are already bounding the mediator mass to be well into the TeV range for WIMP-like scenarios.
This motivates a consistent and systematic exploration of the parameter space to map out possible regions where the rates could be suppressed.
We do indeed find such regions and proceed to construct consistent UV models that generate the relevant effective theory.
We then discuss the corresponding constraints from
both collider and direct-detection experiments on the same parameter space.
We find a benchmark scenario, where even for future XENONnT experiment, LHC constraints will have a greater sensitivity to the mediator mass.
}

\preprint{LTH 1298,~DESY-22-022}

\begin{document}
\maketitle
\flushbottom

\section{Introduction}

While weakly interacting massive particles (WIMPs) remain attractive candidates for explaining the dark-matter (DM) content in the Universe, the null results from the leading direct-detection experiments LUX~\cite{LUX:2016ggv,LUX:2017ree} and XENON1T~\cite{XENON:2018voc,XENON:2019rxp} severely constrain the interaction rates between WIMPs and the Standard Model (SM) particles.
It is possible however to retain sizeable WIMP--quark couplings while suppressing the direct-detection rates by tuning the up and down quark interaction strengths in order to cancel the coherent spin-independent contributions of protons and neutrons in a particular isotope, as was shown e.g. in Ref.~\cite{Kopp:2010su} for the case of vectorial DM couplings.

In this paper, we focus on axial-vectorial DM interactions, where the direct-detection rate is suppressed either due to the absence of coherent enhancement or via dependence on the velocity of the DM in the halo or on the momentum exchange between the DM and the nucleus~\cite{Jungman:1995df}.
The continuously tightening constraints led by XENON1T make this scenario phenomenologically relevant \cite{Blanco:2019hah}, and if the null results persist in the future, further suppression will be required to justify the absence of signals.
However, in the case of an axial-vectorial DM coupling, the suppression by isospin-breaking interactions is
more complicated
since axial-vectorial and vectorial quark currents contribute equally to the scattering cross section as we will show.
Furthermore, the $\mathrm{SU}(2)\times \mathrm{U}(1)$ gauge invariance of the SM implies that the $V-A$ (vectorial minus axial) couplings of up and down quarks have equal strength, i.e. the effective dimension-6 interactions,
\begin{equation}
\label{eq:up-down-v-minus-a}
(\bar\chi \chi)_A(\bar u u)_{V-A} \leftrightarrow (\bar\chi \chi)_A(\bar d d)_{V-A}
\end{equation}
are directly related.
Consequently, it is not clear how efficiently the isospin suppression can occur for axial-vectorial DM currents.

To consistently incorporate the SM gauge invariance, we use the effective theory above the electroweak (EW) scale from Ref.~\cite{Bishara:2018vix} that couples SU(2)$\times$U(1) invariant SM fields with axial-vectorial DM currents.
Such a theory naturally arises in models where the DM candidate is a Weyl
fermion and the coupling between the dark and visible sectors is mediated by a tree-level neutral vector-boson exchange, typically referred to as $Z^\prime$ and often related to an additional U$(1)^\prime$ gauge symmetry that is spontaneously broken at a scale $M_* \gg \mw$. This setup gives rise to a Majorana fermion in the broken phase below $M_*$.
In such models, the vectorial couplings between the DM candidate and the $Z^\prime$ vanish because Majorana fermions are self-conjugate under charge conjugation, while the vectorial current is odd.
In addition, the extra U$(1)^\prime$ gauge symmetry in these models imposes constraints from anomaly cancellation similar to those studied in Ref.~\cite{Ellis:2017tkh} for Dirac DM. The UV completion is also needed to cure
the breakdown of perturbative unitarity above the cutoff of the effective theory~\cite{Kahlhoefer:2015bea,Englert:2016joy,Jacques:2016dqz,Cui:2017juz}.

Experimentally, this scenario can be tested both at colliders and direct-detection experiments.
At the LHC, this is primarily done by searching for an excess in the monojet final state, and projecting the excluded cross section onto the parameter space of simplified benchmark models~\cite{ATLAS:2021kxv, CMS:2021far}, selected based on recommendations from both the theory and experimental communities~\cite{Buchmueller:2014yoa, Abdallah:2015ter, Abercrombie:2015wmb, Boveia:2016mrp}.
The analysis of other final states, especially dijets~\cite{ATLAS:2019fgd, ATLAS:2018qto} and dileptons~\cite{ATLAS:2015rbx,CMS:2016xbv,ATLAS:2019erb,CMS:2020ulv,CMS:2021ctt}, can lead to even tighter, albeit more model-dependent, constraints on the couplings and mass of the $Z^\prime$ mediator.
The non-observation of signals in these searches generally requires mediator masses around the TeV scale.

In direct-detection experiments, the large separation between the $Z^\prime$ mass and the momentum transferred in the scattering of DM off nuclei justifies the adoption of an EFT description of the interaction between the DM and the baryons and mesons.

This paper is organised as follows. In Sec.~\ref{sec:dmeft} we identify the relevant operators for the EFT of axial-vectorial DM current and study the possible suppression of direct-detection rates. In Sec.~\ref{sec:uv-considerations} we consider the constraints arising from the embedding of the low-energy theory into a consistent UV completion, with particular focus on the gauge-anomaly-cancellation requirement. We then identify minimal UV-consistent benchmark models featuring near maximal direct-detection suppression.
In Sec.~\ref{sec:technical_details} we outline the technical details for the calculation of the experimental constraints, and in Sec.~\ref{sec:results} we compare the sensitivities of current and future direct-detection experiments against the ones from collider searches within the selected benchmark scenarios.
Finally, we conclude in Sec.~\ref{sec:conc}. We refer to Appendix~\ref{app:anomalies} for details about the anomaly cancellation and Appendix~\ref{sec:uv-model} outlines an example UV-complete model and
details how the UV model with a single Weyl fermion results in low-energy Majorana DM model with axial-vectorial couplings.

\section{Effective Field Theory for Direct Detection of Axial Vector Dark Matter}
\label{sec:dmeft}

We consider an axial-vectorial DM current coupled to the SM at the dimension-6 level in the unbroken EW phase. The resulting
dark-matter-EFT (DMEFT) Lagrangian can be written as~\cite{Bishara:2018vix},
\begin{equation}
\mathcal{L}_\textsc{dmeft} = \sum_{i,d} \crmn{i}{d}Q^{(d)}_i\equiv \frac{\hat{c}_i^{(d)}}{\Lambda^{d-4}}Q^{(d)}_i\xrightarrow{\textrm{EWSB}}
\sum_{i,d}\cscr{i}{d}\mathcal{Q}_i^{(d)}\,,
\end{equation}
where we used a curly-script notation for the operators and their coefficients below the EW scale to distinguish them from the ones above it. The lower-case hatted coefficients are dimensionless (in natural units) while the upper-case un-hatted ones are dimensionful.

To be concrete, if $\chi$ is an $\textrm{SU}(2)$-singlet Majorana fermion,
the following three operators coupling $\chi$ to the SM quarks will be generated above the EW scale (following the notation of Ref.~\cite{Bishara:2018vix}),
\begin{equation}
Q_{6,i}^{(6)} = (\bar\chi\gamma_\mu \gamma_5 \chi)(\bar Q_L^i \gamma^\mu Q_L^i)\,,\quad
Q_{7,i}^{(6)} = (\bar\chi\gamma_\mu \gamma_5 \chi)(\bar u_R^i \gamma^\mu u_R^i)\,,\quad
Q_{8,i}^{(6)} = (\bar\chi\gamma_\mu \gamma_5 \chi)(\bar d_R^i \gamma^\mu d_R^i)\,,
\label{eq:dim6:Q48}
\end{equation}
where $i=1,2,3$ denotes the quark generation.
In principle, we should also include the operator involving the Higgs-current, $iH^\dagger \stackrel{\leftrightarrow}{D}_\mu H$, since it mixes with the operators in Eq.~\eqref{eq:dim6:Q48} above the EW scale and matches onto
both operators in Eq.~\eqref{eq:dim6EW:Q3Q4:light}
%with both vectorial and axial-vectorial quark-currents
below it~\cite{Bishara:2018vix,Crivellin:2014qxa,DEramo:2014nmf,Brod:2018ust}.
However, this effect is not relevant in our setup because both quark currents appear anyway. Furthermore, the vectorial quark-current does not lead to an enhanced direct-detection rate since we are only concerned with the axial-vectorial DM current.
After EW symmetry breaking, the three operators in Eq.~\eqref{eq:dim6:Q48} match onto the following two:
\begin{align}
{\cal Q}_{2,q}^{(6)} & = (\bar \chi \gamma_\mu\gamma_5 \chi) (\bar q \gamma^\mu q)\,,
& {\cal Q}_{4,q}^{(6)}& = (\bar
\chi\gamma_\mu\gamma_5 \chi)(\bar q \gamma^\mu \gamma_5 q)\,.
\label{eq:dim6EW:Q3Q4:light}
\end{align}
The matching conditions for operators involving first-generation quarks are,
\begin{equation}
\begin{split}
(A\otimes V)_u:\quad \cscr{2,u}{6} &= \crmn{7,1}{6}+ \crmn{6,1}{6}\,,\qquad\qquad
(A\otimes V)_d:\quad \cscr{2,d}{6} = \crmn{8,1}{6}+ \crmn{6,1}{6}\,,\\
(A\otimes A)_u:\quad \cscr{4,u}{6} &= \crmn{7,1}{6}- \crmn{6,1}{6}\,,\qquad\qquad
(A\otimes A)_d:\quad \cscr{4,d}{6} = \crmn{8,1}{6}- \crmn{6,1}{6}\,,
\end{split}\label{eq:match6>5}
\end{equation}
where the shorthand notation on the left-hand side of the colon gives the Lorentz structure of the operator as a product of DM and SM currents, respectively, $(A) V$ stands for the (axial-)vectorial current, and the subscript denotes the quark flavour of the operator.

Unlike in the spin-independent case, however, if the coupling to the DM is purely axial, both the $A\otimes V$ and $A\otimes A$ operators can contribute equally to the direct-detection cross section for heavy nuclei (i.e. with mass number $\mathscr{A}\gtrsim 100$). Consequently, it is not trivial to obtain a similar suppression as in the spin-independent case.
It is already known that in the $V\otimes V$ interaction, the cancellation can be effected by tuning the non-relativistic coefficients of the proton and neutron operator, i.e. by breaking isospin symmetry. For a single isotope, this cancellation can be complete only in the zero momentum transfer limit, $q\to 0$.
For the interactions of concern here, the mechanism is the same. The complication arises when both vectorial and axial SM currents contribute equally to the cross section. The reason is that the amount, and more importantly the sign, of the isospin-breaking ratio required to cancel each interaction is different.
This is because the vectorial currents count the number of up and down quarks in the nucleus, therefore, the coefficients of the operators containing up and down currents have to have opposite signs in order for the interference term to be negative.
For the axial current, on the other hand, the contribution of the up and down quarks to the spin of the nucleon have opposite signs to begin with. Thus, the coefficients of the operators containing axial up and down currents have to have the same sign for the interference term to be, again, negative.
The ability to align the minima of both the $\axa$ and $\axv$ stems from the fact that the former is much more sensitive to which combination of the two right-handed quark currents (up vs. down) implements isospin breaking.

\begin{figure}[t]\centering
    \begin{subfigure}[t]{0.48\textwidth}\centering
		\includegraphics[width=\textwidth]{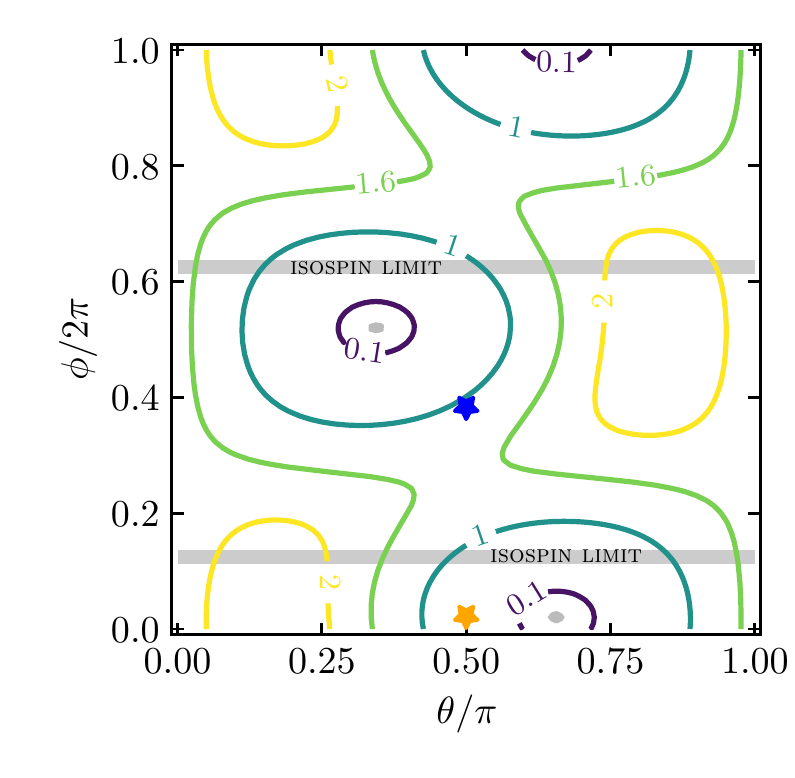}
		\caption{Contours of the relative direct-detection event rate in the $\theta-\phi$ plane using the parameterisation of Eq.~\eqref{eq:spherical}. The thick horizontal lines correspond to the isospin limit.}
		\label{fig:contours0}
      \end{subfigure}\hfill
    \begin{subfigure}[t]{0.48\textwidth}\centering
    \includegraphics[width=\textwidth]{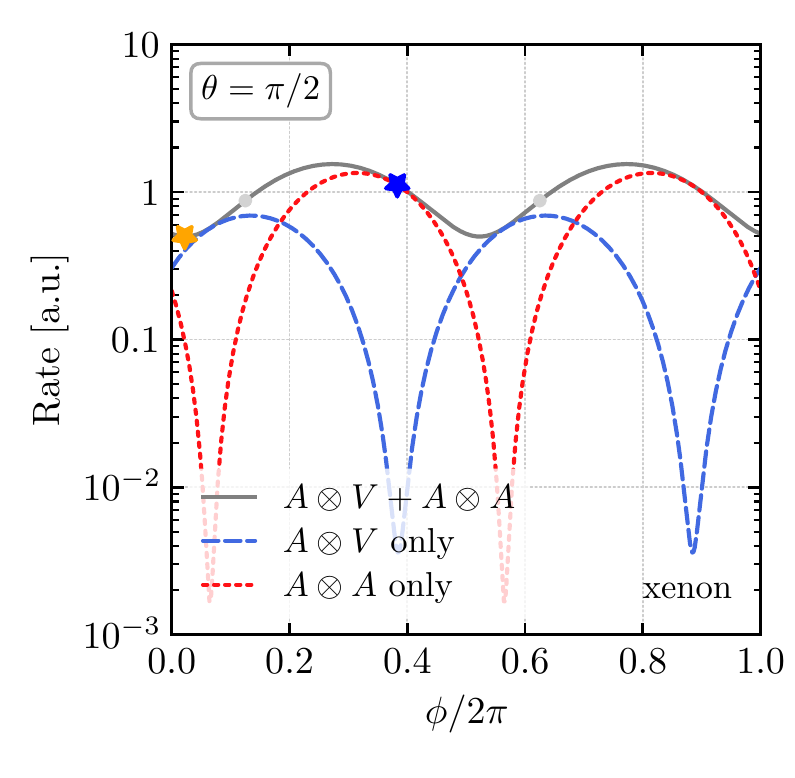}
    \caption{Relative event rate along $\theta=\pi/2$ corresponding to $C_{6,1}^{(6)}=0$. }
    \label{fig:rate:theta=pi/2}
  \end{subfigure}\\
  \begin{subfigure}{0.48\textwidth}\centering
  \includegraphics[width=\textwidth]{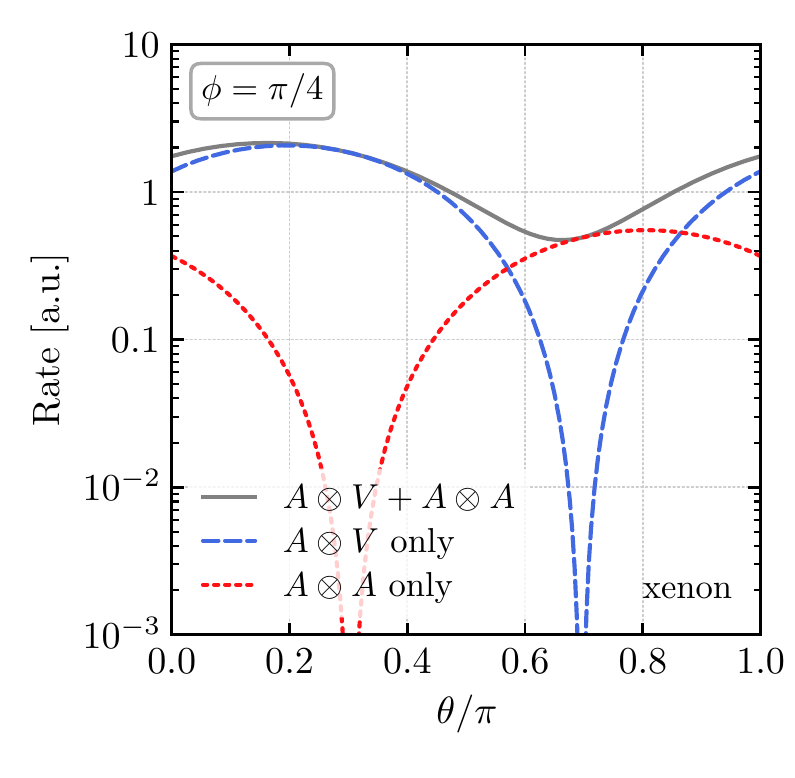}
  \caption{Relative event rate along $\phi=\pi/4$, i.e. the isospin limit. }
  \label{fig:rate:isospin_limit}
\end{subfigure}\hfill
\begin{subfigure}{0.48\textwidth}\centering
\includegraphics[width=\textwidth]{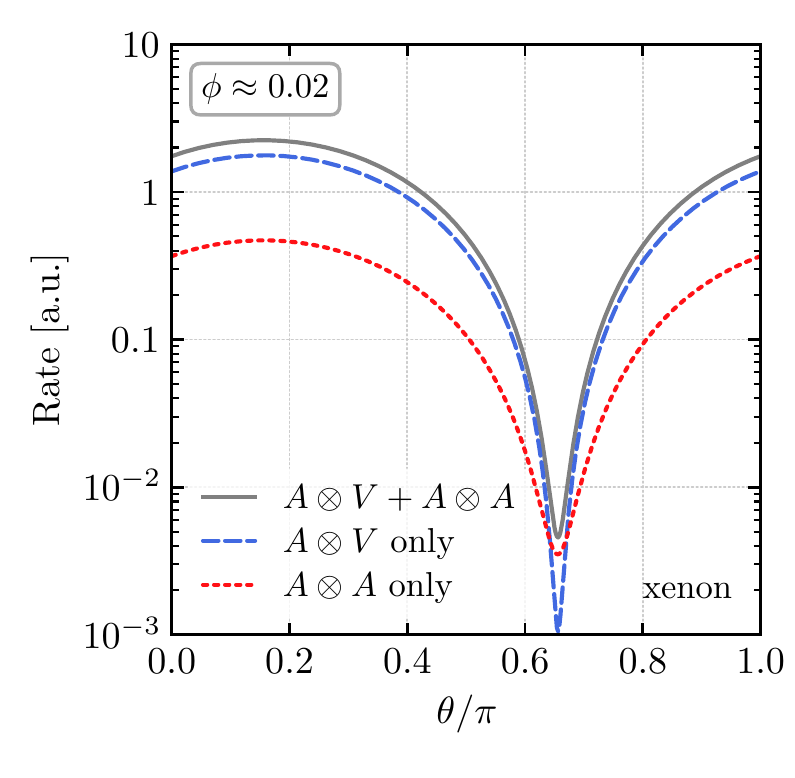}
\caption{Relative event rate along $\phi=0.02$ that passes throught the global minimum. }
\label{fig:rate:global_min}
\end{subfigure}
  \caption{The direct detection rate for $m_\chi=100$ GeV using the parameterisation of Eq.~\eqref{eq:spherical} with normalisation corresponding to the overall size of the Wilson coefficients. The blue (orange) star shows the benchmark point BM1 (BM2). The solid grey curve in the upper-right panel, (\subref{fig:rate:theta=pi/2}), and both lower panels, (\subref{fig:rate:isospin_limit})~\&~(\subref{fig:rate:global_min}), show the combined relative rate with the normalisation of panel~(\subref{fig:contours}) along different lines in the two-dimensional plot, while the dashed (blue) and dotted (red) curves show the sole $A\otimes V$ and $A\otimes A$ contributions along the same line, respectively.
    }
  \label{fig:rel_rates0}
\end{figure}

The matching conditions in Eq.~\eqref{eq:match6>5} present us with different possibilities as follows.
\begin{enumerate}
	\item We can entirely eliminate either the vectorial, $V$, or axial-vectorial, $A$, currents on the SM side (i.e. for all flavours). This can be accomplished with the choice,
	\begin{equation}
	\crmn{6,1}{6} = \mp\crmn{7,1}{6} = \mp\crmn{8,1}{6}\,,
	\end{equation}
	which automatically enforces isospin-symmetric coefficients in the EFT.
	\item If isospin breaking is desired, neither the vectorial nor the axial-vectorial currents can be completely eliminated, and one is left with a mixture of both, see Eq.~\eqref{eq:match6>5}. As we will show, this makes suppressing the direct detection rate more complicated since only special regions in parameter space allow for the simultaneous suppression of the contributions from both currents\,--\,see, e.g., Figs.~\ref{fig:rate:theta=pi/2}\,--\,~\ref{fig:rate:global_min}.
\end{enumerate}

To simplify the analysis, we adopt the following parametrisation for the coefficients,
\begin{equation}
C_{6,1}^{(6)}\to \frac{g^{\prime\,2}}{\Lambda^2}\cos\theta\,,\qquad
C_{7,1}^{(6)}\to \frac{g^{\prime\,2}}{\Lambda^2}\sin\theta\cos\phi\,,\qquad
C_{8,1}^{(6)}\to \frac{g^{\prime\,2}}{\Lambda^2}\sin\theta\sin\phi\,,
\label{eq:spherical}
\end{equation}
such that the sum of their squares equals $g^{\prime\,4}/\Lambda^4$. This parametrisation removes one variable and allows us to project the relative direct-detection rate, in a model with only first-generation quarks coupled to DM, as function of the overall value of the Wilson coefficients in the $\theta-\phi$ plane.

The contours in Fig.~\ref{fig:contours0} represent the relative direct-detection rates for a reference DM mass $m_\chi$ = 100 GeV.
The figure shows that for specific choices of the parameters a factor of roughly $10^{-2}$ suppression can be achieved. For heavier DM masses, while the overall event rate falls as $1/m_\chi$, there is no qualitative change in the arguments presented here.
The isospin-symmetric limit is realized on two horizontal lines in the $\theta-\phi$ plane with $\phi=\pi/4,5\pi/4$. These lines are shown in Fig.~\ref{fig:contours0} as thick grey lines labelled `\textsc{isospin limit}'.

Figure~\ref{fig:rate:theta=pi/2} shows the relative rate with the normalisation of Fig.~\ref{fig:contours0} along the curve $\theta=\pi/2$ corresponding to scenarios where the coefficient $C_{6,1}^{(6)}=0$, thus where only first-generation right-handed quarks are coupled to the DM current. The $A\otimes V$ and $A\otimes A$ contributions are shown separately in dashed blue and dotted red curves, respectively, and the combination of these two, i.e. the total relative rate corresponding to the contours in Fig.~\ref{fig:contours0}, in solid gray.
The gray dots denote the isospin limit while the blue and orange stars, also reported in Fig.~\ref{fig:rate:theta=pi/2}, correspond to the benchmark scenarios BM1 and BM2 that will be analysed in Sec.~\ref{sec:results}.
The scenario BM1 corresponds to the case where the $A\otimes V$ rate is significantly suppressed, and naive treatment without considering also the $A\otimes A$ contribution would lead to a wrong conclusion since, in fact, only $\mathcal{O}(1)$ suppression of the direct-detection rate can be obtained.
The benchmark BM2 has instead the minimal direct-detection rate in this construction where the first-generation doublet does not couple to the DM current.

Figure~\ref{fig:rate:isospin_limit} depicts the relative rate with the same normalisation in the isospin limit showing explicitly that significant suppression is not possible by cancelling separately the $A\otimes V$ or the $A\otimes A$ contribution.
Finally, Fig.~\ref{fig:rate:global_min} illustrates the rate along the $\phi\approx 0.02$ line passing through the global minimum in the $\theta-\phi$ plane showing that the largest suppression can be achieved in the parameter space point where both $A\otimes V$ and $A\otimes A$ simultaneously have a minimum.

\section{UV completions and anomaly cancellation}
\label{sec:uv-considerations}

A straightforward path to UV completing the EFT setup in Sec.~\ref{sec:dmeft} is to augment the SM gauge group by a $\textrm{U}(1)^\prime$ group which couples to the SM fermions and the DM candidate via a heavy $Z^\prime$, $m_{Z^\prime}\gg m_Z$.
For the $Z^\prime$ to couple axial-vectorially to the DM, the latter has to be chiral under the $\textrm{U}(1)^\prime$.
Apart from the need to Higgs this gauge symmetry in order to generate a mass for the mediator, the DM and SM fermion $\textrm{U}(1)^\prime$ charges must be chosen such that pure and mixed gauge anomalies cancel.
In the following, we briefly discuss this and other general aspects that arise from considering possible UV completions of the EFT setup, and we refer to Appendix~\ref{app:anomalies} for general anomaly equations involving first- and second-generation SM quarks, and to Appendix~\ref{sec:uv-model} for an explicit construction of a one-generation model that addresses (most) of these general points.

\begin{enumerate}[(i)]
\item \emph{Anomaly cancellation.}
Anomaly-free DM models were discussed in Ref.~\cite{Ellis:2017tkh,Cui:2017juz,FileviezPerez:2019jju,Perez:2020jyg,Costa:2019zzy} where, however, the minimal model
with the additional matter-field content consisting of only one Weyl fermion was not discussed.
The general anomaly equations that must be satisfied are given in Appendix~\ref{app:anomalies} and their solution in this case requires that the SM fermions to be charged under $\textrm{U}(1)^\prime$.

\item \emph{Couplings of the $Z^\prime$ to leptons.} A feature of the mixed anomaly equations is that charges of the SM fermions are, in general, a linear combination of their hypercharge $Y$ and $B-L$ where $B(L)$ are the baryon(lepton) numbers which are $\pm\tfrac{1}{3}(\pm 1)$.
This general statement has a significant consequence, namely that coupling the $Z^\prime$ to the SM leptons is unavoidable, and thus observables involving leptons must be taken into account.

\item \emph{SM Yukawa couplings.} The $\textrm{U}(1)^\prime$ gauge invariance forbids some SM Yukawa couplings. We show a possible mechanism for generating the Yukawa couplings in the model with only one generation of SM quarks carrying U$(1)'$ charge in Appendix~\ref{sec:uv-model}. This construction includes a dark Higgs, $S$, charged under $\textrm{U}(1)^\prime$ and generating the effective Yukawa couplings upon spontaneous breaking of the symmetry.
Another option is to add Higgs doublets that also carry a $\textrm{U}(1)^\prime$ charge \cite{Ko:2012hd}.

\item \emph{Tree-level-induced spin-independent contributions.}
Depending on the scalar sector of the theory, the tree-level exchange of  physical scalars could induce a sizeable spin-independent cross section. However, we stress that this is model dependent and could be suppressed, for example, in the following two ways.
First, with one dark Higgs, a small U$(1)'$ gauge coupling allows the mass of the scalar to be significantly above that of the $Z'$. Second, it could be tuned to zero at the tree level by extending the scalar sector such that no single state couples to both the quarks and the DM sector at the same time.
In this case, the mixing between the scalars would still generate this interaction though it would be additionally suppressed by at least a loop factor.

\item \emph{Loop-induced spin-independent contributions.}
While spin-independent scattering is suppressed at the tree level, it can be induced at the one-loop level depending on the UV completion via either two insertions of the axial-vectorial coupling or by a potential dark Higgs penguin with a $Z'$ in the loop.
Using naive dimensional analysis, we estimate the relative size of the loop-induced scalar-scalar interaction via internal vector-boson exchange in comparison with the tree-level current-current interaction:
\begin{equation}
\frac{\sigma_\textsc{si}}{\sigma_\textsc{sd}}\sim A^2\frac{g^{\prime\,4}}{(4\pi)^4}\frac{m_N^2m\chi^2}{m_{Z^\prime}^4}\sim \mathcal{O}(10^{-11})\,,
\end{equation}
where we used $g^\prime=0.1$, $m_{Z^\prime}=1$ TeV, $m_\chi = m_{Z^\prime}/2$, and $m_N=1$ GeV is the nucleon mass. We also took the mass number of the atomic nucleus (xenon, for example) to be $\mathscr{A}=100$. The dependence on the DM and nucleon masses arises from the required chirality flips in the scalar-scalar operator. Whether one should insert the nucleon mass or $\Lambda_\textsc{qcd}$ is irrelevant to the estimate, but we note that the respective form factors would suppress the spin-independent contribution even further. The heavy suppression of the spin-independent scattering rate and its model-dependent origin justifies our choice of neglecting this contribution

\item \emph{Flavour violation.}
In models where only the right-handed quarks are charged under the U$(1)'$, the mass- and gauge-bases can always be aligned using the freedom in choosing the right-handed rotation matrices.
However, in models where the three generations of left-handed quark doublets carry non-universal U$(1)'$ charges, and working the down-basis, flavour violation in the up quark sector is unavoidable. As a result, the $Z'$ gauge boson mediates tree-level flavour-changing neutral currents. Thus, neutral $D$ meson mixing provides strong constraints on the breaking of the first- and second-generation U$(2)$ flavour symmetry which bounds the $Z'$ mass to be $m_{Z'}\gtrsim 50$~TeV for the model in Table~\ref{tab:1g+sr_model}, see Ref.~\cite{UTfit:2007eik}. Retaining the U$(2)$ symmetry can lower this bound significantly.
\end{enumerate}

\begin{figure}[t]\centering
	\begin{subfigure}[t]{0.48\textwidth}\centering
		\includegraphics[width=\textwidth]{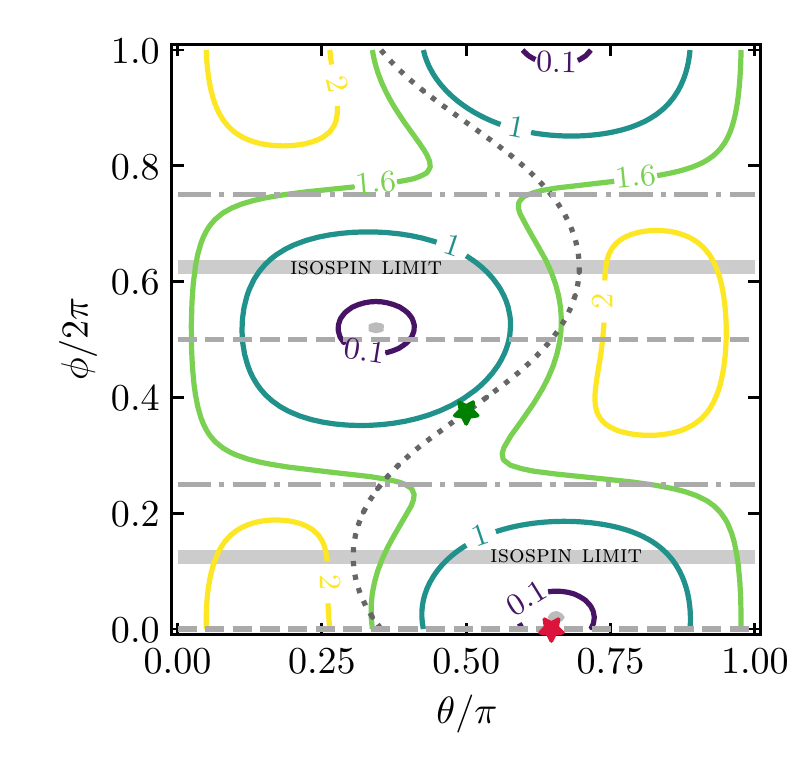}
		\caption{The dotted, dashed, and dash-dotted curves correspond to gauge-anomaly-free scenarios S1, S2 and S3, respectively. The thick horizontal lines correspond to the isospin limit.}
		\label{fig:contours}
	\end{subfigure}
  \hfill
  \begin{subfigure}[t]{0.48\textwidth}\centering
  	\includegraphics[width=\textwidth]{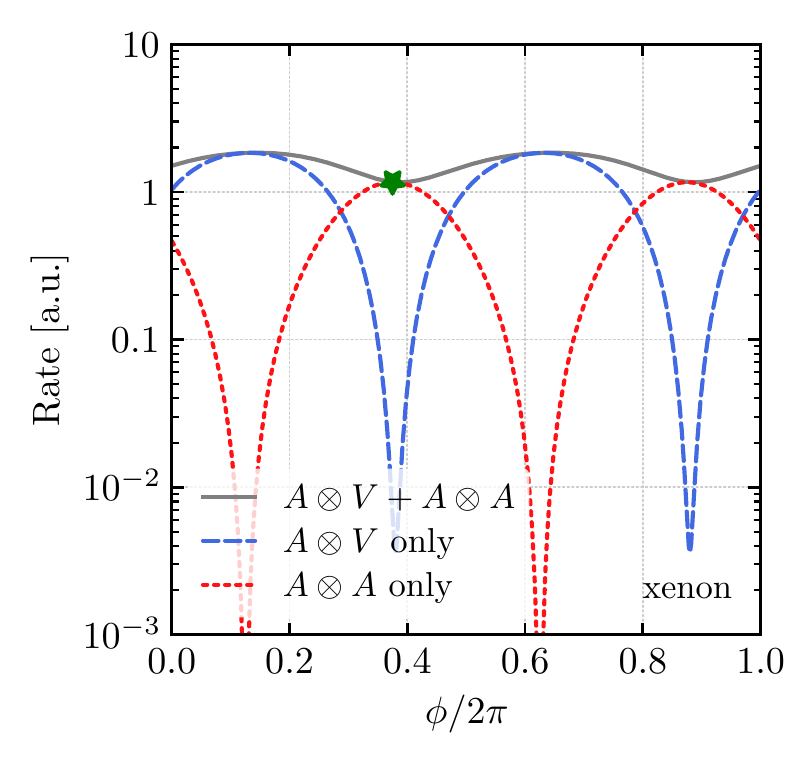}
  	\caption{Relative event rate along $\tan\theta=2/(\sin\phi+\cos\phi)$.}
  	\label{fig:rate:theta,phi}
  \end{subfigure}
	\centering
	\begin{subfigure}{0.49\textwidth}
		\includegraphics[width=\textwidth]{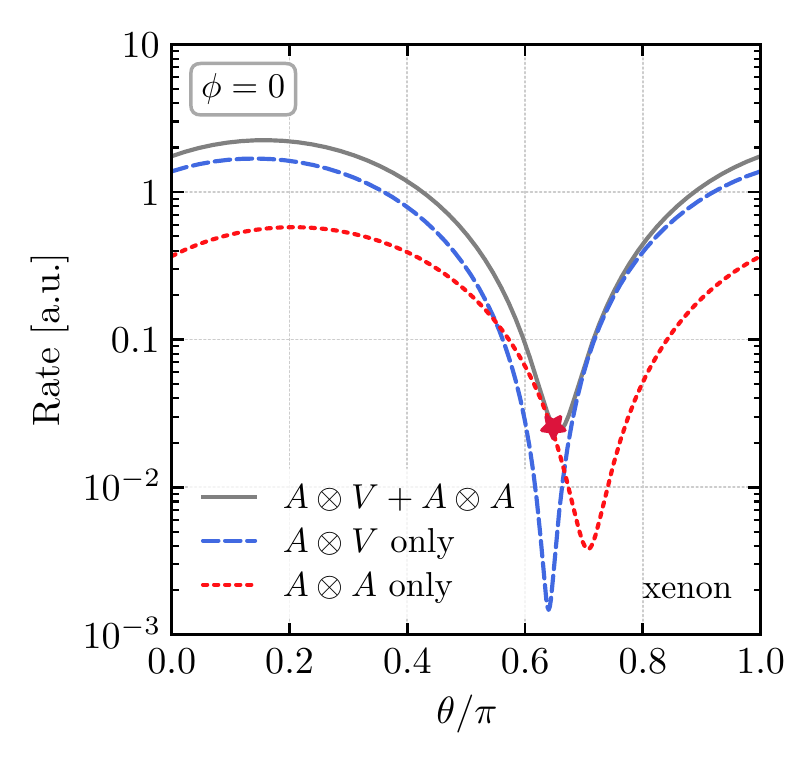}
		\caption{Relative event rate along $\phi=0$.}
		\label{fig:rate:phi=0}
	\end{subfigure}\\

	\caption{The direct-detection rate for $m_\chi=100$ GeV using the parameterisation of Eq.~\eqref{eq:spherical} with normalisation corresponding to the overall size of the Wilson coefficients. The green (red) star shows the benchmark point BM3 (BM4)in Table~\ref{tab:1g+sr_cr_model}. The solid grey curve in upper right and lower panels shows the combined relative rate with the normalisation of panel~(\subref{fig:contours})  along $\phi=0$, while the dashed (blue) and dotted (red) curves show the sole $A\otimes V$ and $A\otimes A$ contributions along the same curve, respectively.
	}
	\label{fig:isospin-breaking}
\end{figure}

We use the parameterisation of Eq.~\eqref{eq:spherical} to study the potential suppression of the direct-detection rate in minimal anomaly-free UV completions of the scenario.
With only one dark Weyl fermion and only first-generation quarks charged under U$(1)^\prime$ (or in the case of equal charges for all generations), we are restricted to lie on the dotted gray line of Fig.~\ref{fig:contours}, which corresponds to the choice $\tan\theta=2/(\sin\phi+\cos\phi)$.
The projection of the direct-detection rate with the same normalisation as before is shown in Fig.~\ref{fig:rate:theta,phi}.
Here the green star represents the choice with the minimal relative rate for this one-generation scenario, confirming that only $\mathcal{O}(1)$ suppression is achievable.
A more detailed discussion about the construction of the corresponding UV complete model of Eq.~\eqref{eq:charges} is given in Appendix~\ref{sec:uv-model}.

To reach near maximal suppression in the $\theta-\phi$ plane, we need to augment the minimal one-generation scenario with generation-dependent charges. The dot-dashed (dashed) horizontal gray lines in Fig.~\ref{fig:contours} show the anomaly-free solutions with second-generation up(down)-type quark carrying
an independent U$(1)^\prime$ charge, respectively. This solution is referred to as S2 (S3) in  Appendix~\ref{app:anomalies}. We checked explicitly that the contribution of the strange-quark current to the direct detection is sub-leading and does not affect the positions of the minima as expected.

Figure~\ref{fig:rate:phi=0} shows the relative event rate using the same arbitrary normalization factor along the $\phi=0$ curve.
Along this projection, we identify the BM3 (red star) whose direct-detection event rate is very close to the achievable minimum, and which will be analysed in Sec.~\ref{sec:results}.
This model is realised by charging the first-generation left- and right-handed quarks and the right-handed strange quark under U$(1)^\prime$, while we choose to couple only the third-generation leptons in order to alleviate the stringent bounds from dilepton searches.
The specific charge assignments for the BM3 benchmark are summarised in Table~\ref{tab:1g+sr_model}.

\begingroup
\setlength{\tabcolsep}{12pt}
\begin{table}\centering
	\begin{tabular}{c c c c c c c c c}\toprule[1pt]
		& $Q_{L,1}^c$ & $u_{R,1}$ & $d_{R,1}$ & $u_{R,2}$ & $d_{R,2}$ & $L_{L,3}^c$ & $e_{R,3}$ & $\chi_R$\\\midrule[0.5pt]
    BM3 & $+\frac{1}{2}$ & +1 & 0 & 0 & -2 & $-\frac{3}{2}$ & 0 & 3 \\
		\bottomrule[1pt]
	\end{tabular}
	\caption{The benchmark model BM3 with only one addtional Weyl fermion and first-generation quarks, right-handed second-generation down quarks and third-generation left-handed leptons carrying U$(1)^\prime$ charge.}
	\label{tab:1g+sr_model}
\end{table}
\endgroup

In the next sections, we study the experimental constraints for the benchmark models, and compare the
current and future sensitivities at direct-detection experiments against up-to-date collider exclusions.

\section{Technical details for experimental constraints}
\label{sec:technical_details}
\subsection{Direct detection}

The DM direct-detection experiments search for signals from DM scattering off atomic nuclei in shielded detectors.
For concreteness, we consider the XENON1T experiment with an exposure of  278.8 days $\times$ 1300 kg~\cite{XENON:2018voc}, and also the projected exposure of 20 ton $\times$ year for the XENONnT \cite{XENON:2020kmp}.
The scattering rate $\mathcal R$, which is
the expected number of events per detector mass per unit time, can be  expressed differentially with respect to the recoil energy as~\cite{Lewin:1995rx},
\begin{equation}\label{eq:rate}
\frac{d\mathcal R}{dE_{\rm R}} = \frac{\rho_\chi}{m_{A}m_\chi}\int_{v_\text{min}}
\frac{d\sigma}{dE_{\rm R}}vf_\oplus(\vec v)d^3\vec v,
\end{equation}
where $E_{\rm R}$ is the recoil energy of the nucleus, $m_A$ is the mass of the nucleus, and $\rho_\chi$ is the local DM density.
We approximate the DM velocity in the halo, $f_\oplus(\vec v)$, with a Boltzmann distribution and integrate over recoil energies in the range $E_R\in[3,40]$ keV~\cite{XENON:2018voc} to approximate the detector efficiency.
The exclusion curves in Fig.~\ref{fig:axial_Wilson} were obtained (naively) using Poisson statistics assuming zero events in the signal region.
The coefficients of the Galilean-invariant effective theory~\cite{Fitzpatrick:2012ix} were computed using \texttt{DirectDM}~\cite{bishara2017directdm} and the nuclear responses and direct-detection rates were obtained with \texttt{DMFormFactor}~\cite{anand2013model}.

\subsection{Monojet searches at the LHC}

Model-independent searches for DM at the LHC are primarily performed via monojet~\cite{ATLAS:2021kxv, CMS:2021far} and monophoton analyses~\cite{ATLAS:2020uiq, CMS:2017qyo}, which can be interpreted in the context of simplified DM models, see Refs.~\cite{Buchmueller:2014yoa, Abdallah:2015ter, Abercrombie:2015wmb}, with the monojet final state typically providing the stronger limits.
Here we consider the recent DM analysis by the ATLAS collaboration~\cite{ATLAS:2021kxv} which uses 139 fb$^{-1}$ data and where monojet events with large missing energy were used to constrain a simplified model with a vector mediator, $Z^\prime$, coupling to the DM, $\chi$, and the quark, $q$, excluding $Z^\prime$ with masses around 2~TeV.
The quoted limits on axial mediators are very similar.

We were able to reproduce the exclusion limits with good approximation for the simplified DM benchmark model adopted in the experimental analyses~\cite{Buchmueller:2014yoa, Abdallah:2015ter, Abercrombie:2015wmb, Boveia:2016mrp}, where the couplings of the vector mediator $Z^\prime$ to quarks and the DM are fixed to $g_q$ = 0.25 and $g_\chi$ = 1.
We scanned over several configurations of DM and mediator masses by importing the simplified DM model from Ref.~\cite{Backovic:2015soa} into \textsc{MadGraph5\_aMC@NLO}~\cite{Alwall:2014hca} to generate the WIMP $s$-channel process $pp \to j V \to j \bar\chi\chi$ where $j$ is a jet from initial state radiation, and the DM particle pair, $\bar\chi\chi$, gives rise to missing transverse energy, $E_T^{\rm miss}$, in the detector.
The process is implemented at LO in the strong coupling constant. We adopted the NNPDF3.0\_LO PDF set~\cite{NNPDF:2014otw}, and for each event the factorization and renormalisation scales were set to $H_T / 2$, with the total hadronic transverse energy $H_T = \sqrt{m_{\chi\chi}^2 + p_{T,j}^2} + p_{T,j}$ where $m_{\chi\chi}$ is the invariant mass of the DM pair, and $p_{T,j}$ is the transverse momentum of the parton-level jet.
Events are hadronised using \textsc{Pythia8}~\cite{Sjostrand:2007gs}, and a fast detector simulation is carried out using \textsc{Delphes}~\cite{deFavereau:2013fsa}.
We apply the kinematic cuts from Ref.~\cite{ATLAS:2021kxv}, which are as follows:
$E_T^{\textrm{miss}} >$ 200 GeV; leading jet with $p_T >$ 150 GeV and $|\eta| <$ 2.4; no more than three additional jets with $p_T >$ 30 GeV and $|\eta| <$ 2.8; separation between missing transverse momentum and each of the jets $\Delta\phi(\textrm{jet},p_T^{\textrm{miss}}) >$ 0.4 (0.6) for events with $E_T^{\textrm{miss}} >$ 250 GeV (200 GeV $< E_T^{\textrm{miss}} <$ 250 GeV).
The remaining simulated events were binned in thirteen exclusive signal regions as in Ref.~\cite{ATLAS:2021kxv} according to their missing transverse energy.
We simulated a sufficient number of events such that, after the selection cuts, we still obtain a statistically significant sample in all the bins.
Finally we excluded parameter space points where the fiducial cross section of the signal in any bin is bigger than its uncertainty at 95\% confidence, which is evaluated by adding the total systematic uncertainty of the signal quadratically to the statistical uncertainty of the signal and the overall uncertainty of the background (statistical and systematic\footnote{The systematic uncertainty of the signal is obtained by combining the relative uncertainties from Ref.~\cite{ATLAS:2021kxv}: luminosity uncertainty 1.7\%; cross section scale uncertainty 10\%; a PDF uncertainty 5\%; PDF choice 10\%; 1\% to 7\% for the jet $E_T^{\textrm{miss}}$ reconstruction, energy scale and resolution; modelling initial and final state radiation 3\% to 6\%. Scale uncertainty of the signal is neglected. The systematic uncertainties are added linearly, overall systematic and  statistical uncertainties are added in quadrature.}) from Ref.~\cite{ATLAS:2021kxv}.

We validated our analysis by reproducing the existing exclusion limits and by comparing with results from Ref.~\cite{Lozano:2021zbu}, we explored the current LHC monojet sensitivity in the three BMs introduced in the previous sections.
The exclusion limits we obtain will be presented and discussed in Sec.~\ref{sec:results}.

\subsection{Dijets and dileptons}
\label{sec:dijet}

Heavy mediators that couple to quarks can be detected at the LHC via their decays into quarks and leptons. The most recent analysis by the ATLAS collaboration searching for heavy resonances in dijet final states uses 139 fb$^{-1}$ data for state-of-the art constraints~\cite{ATLAS:2019fgd} for mediator masses above 2 TeV. For lower masses above 700 GeV we use the results presented in Ref.~\cite{ATLAS:2018qto} that are based on 29.3 fb$^{-1}$ of data.
For the decay into taus we used the constraints on the cross section of the combined hadronic and leptonic channels~\cite{ATLAS:2015rbx}, which cover mediator masses between 0.5 and 2.5 TeV and are based on an integrated luminosity of $19.5-20.3$ fb$^{-1}$.
We recast these limits into the considered model parameter space, and estimate a lower limit on the $Z^\prime$ mass of about 2 TeV using the code developed in Ref.~\cite{Bishara:2018sgl} except that we use the NNPDF3.0\_LO PDF set.

% \FloatBarrier

\section{Results}
\label{sec:results}

\begin{figure}\centering
    \begin{subfigure}{0.49\textwidth}\centering
		\includegraphics[width=\textwidth]{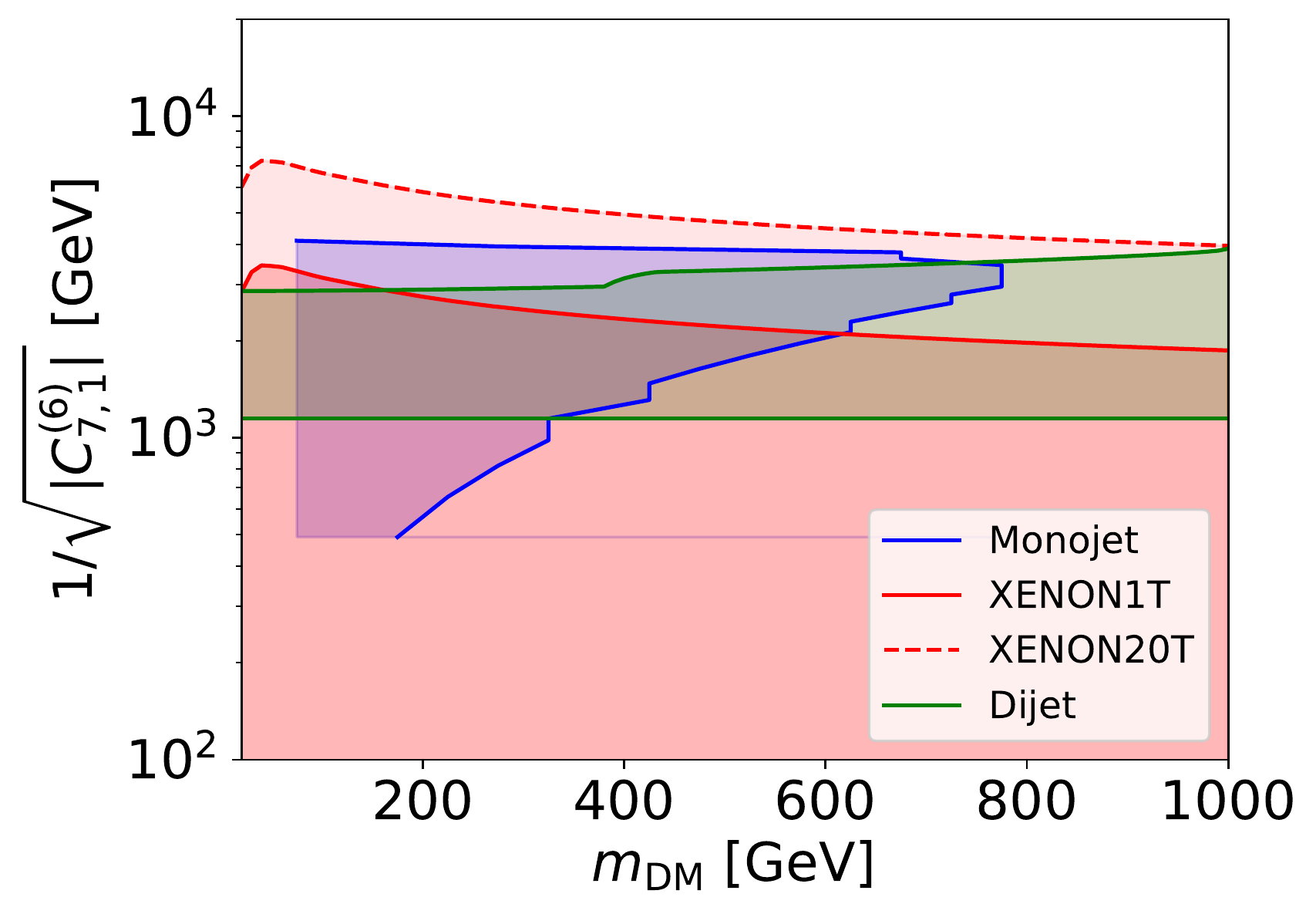}
      \end{subfigure}
    \begin{subfigure}{0.49\textwidth}\centering
    \includegraphics[width=\textwidth]{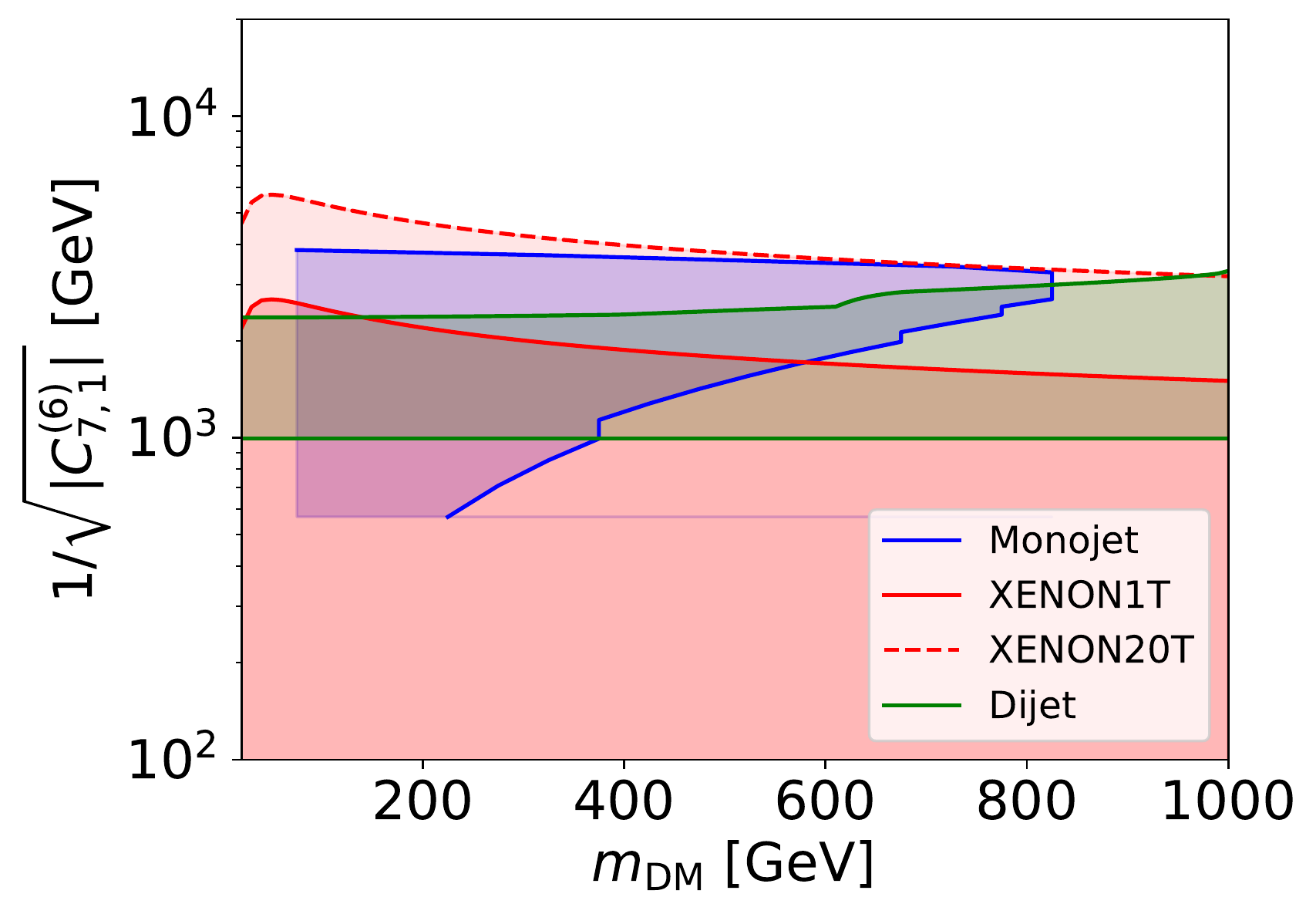}
  \end{subfigure}
  \caption{Exclusion limits for axial mediator couplings to DM from XENON experiment for 1 ton $\times$ year exposure (red solid) and for 20 ton $\times$ year exposure (red dashed), from LHC monojet (blue) and dijet (green) analyses with 139 fb$^{-1}$ integrated luminosity. The results are shown in the plane with the DM mass on the $x$ axis and inverse of the square root of the Wilson coefficient $|C^{(6)}_{7,1}|$ on the $y$ axis for the BM1 (left) and BM2 (right).}
	\label{fig:axial_Wilson}
\end{figure}

In this section we compare the exclusion limits from the XENON experiment and from the LHC  for the introduced BMs.
In Fig.~\ref{fig:axial_Wilson}, we show the results for BM1 and BM2.
The two benchmarks represent realisations of models where only the first generation of quarks is coupled to the DM current through a vector mediator, $Z^\prime$.
They both feature a null charge to the left-handed quarks, i.e. $|C^{(6)}_{6,1}|$ = 0, corresponding to the choice $\theta = \pi / 2$ in the parametrisation of Eq.~\eqref{eq:spherical}.
The exclusion limits from collider monojet and dijet searches and from the current XENON1T and projected XENON20T sensitivities are presented in the plane with the DM mass (in GeV) on the horizontal axis and the (inverse of the square root of the) Wilson coefficient $|C^{(6)}_{7,1}|$ on the vertical axis.
The magnitude of the couplings to right-handed quarks has been fixed in order to match the interaction strength of the experimental benchmarks of Ref.~\cite{Buchmueller:2014yoa, Abdallah:2015ter, Abercrombie:2015wmb, Boveia:2016mrp}, such that a direct comparison with the results in the literature is straightforward.

The benchmark BM1 (left panel of Fig.~\ref{fig:axial_Wilson}) probes the case where the $A\otimes V$ contribution to the direct-detection rate is at minimum (blue star of Fig.~\ref{fig:rate:theta=pi/2}) while the $A\otimes A$ contribution is still substantial.
This case corresponds to the choice $\phi=2.4$ and gives the couplings $g_{u_R} \simeq 0.372$ and $g_{d_R} \simeq -0.334$.
The second benchmark, BM2, (right panel of Fig.~\ref{fig:axial_Wilson}) probes the case where the one-generation model has the minimum overall direct-detection rate ($A\otimes V$ + $A\otimes A$) (orange star of Fig.~\ref{fig:rate:theta=pi/2}).
This case corresponds to the choice $\phi=0.14$ and gives the couplings $g_{u_R} \simeq 0.495$ and $g_{d_R} \simeq 0.069$.

The limits from the monojet analysis (blue curves) depend mainly on the overall magnitude of the interaction, which have been kept fixed in the two BM scenarios in order to be directly comparable with the experimental benchmark~\cite{Buchmueller:2014yoa, Abdallah:2015ter, Abercrombie:2015wmb, Boveia:2016mrp}.
The very marginal differences in the sensitivity curves for our choices of the chiral structure of the couplings arise when the latter are convoluted with the respective PDF weights.
The conservative strongest exclusion of the monojet analyses rules out $Z^\prime$ masses below 1.6 TeV, and it translates into an upper limit reach of about 800~GeV DM, which corresponds to the on-shell limit of the mediator decaying into DM $m_{\rm DM} \leq m_{Z^\prime}/2$.
The limits from dijet searches for mediator masses above 700 GeV (green curves) are largely independent on the DM mass and provide with comparable exclusion to the monojet for light DM.
The rather small increase in sensitivity of the dijet analysis appearing at DM masses of 400 (600) GeV for the BM1 (BM2) occurs at the transition between analyses optimised for intermediate and heavy mediator, as explained in Sec.~\ref{sec:dijet}.
The current XENON sensitivity (solid red curves) is somewhat weaker than the collider exclusions, especially for heavy DM, albeit being able to probe an extended DM mass interval and lighter mediator masses.
On the other hand, the projection for the XENON experiment with 20 ton $\times$ year exposure (dashed red curves) will be able to test  a larger parameter space region in comparison with the current collider reach.

\begin{figure}
	\begin{center}
	\includegraphics[width=0.5\textwidth]{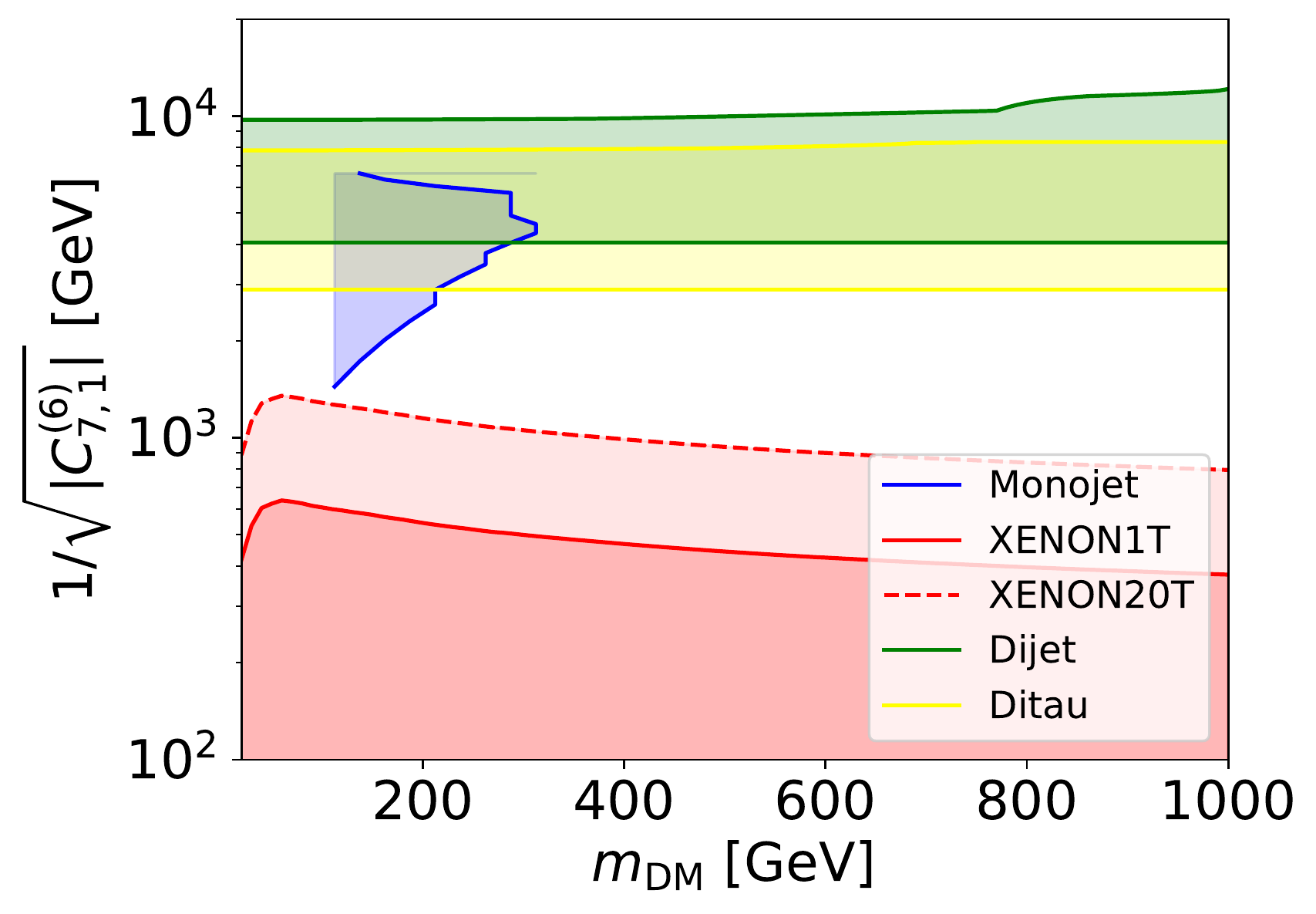}
	\end{center}
	\caption{Exclusion limits for the model of Table~\ref{tab:1g+sr_model} from XENON experiment for 1 ton $\times$ year exposure (red solid) and for 20 ton $\times$ year exposure (red dashed), from LHC monojet (blue) and dijet (green) analyses with 139 fb$^{-1}$ integrated luminosity and ditau (yellow) analyses with $19.5-20.3$ fb$^{-1}$ integrated luminosity.
    The results are shown in the plane with the DM mass on the $x$ axis and the inverse of the square root of the Wilson coefficient $|C^{(6)}_{7,1}|$ on the $y$ axis.}
	\label{fig:strange_bench_Wilson}
\end{figure}

Figure~\ref{fig:strange_bench_Wilson} shows the exclusion limits for the anomaly-free BM3.
The interaction strengths of this model are given by the U$(1)^\prime$ charges of SM and DM particles, reported in Table~\ref{tab:1g+sr_model}, multiplied by an overall gauge coupling that we fix to $g^\prime = 0.1$.
As already mentioned, this construction features an almost minimal direct-detection rate, thus the constraints on heavy mediators from the XENON experiment, both current (solid red line) and future (dashed red line), are outperformed by the collider sensitivity.
Smaller mediator masses, between 300 and 700 GeV, are constrained by neither dijet nor direct-detection searches, see also \cite{Blanco:2019hah}.
Because of the smaller couplings to quarks and DM, the monojet searches (blue curve) can probe only up to 1.2 TeV $Z^\prime$ masses.
As the complete model features other decay channels for the mediator that increase its decay width, the reach of the monojet analysis in terms of DM mass is reduced, and it can only test masses up to about 300~GeV, well below the on-shell limit.
On the other hand the additional decay channels lead to other testable signatures and they can be searched for via dijets (green line) and ditaus (yellow line), which have a stronger sensitivity compared to the monojet searches and are mostly independent of the DM mass.

% \FloatBarrier

\section{Summary and Outlook}
\label{sec:conc}

In this work we considered a WIMP-like Majorana dark matter candidate that mainly interacts through an axial-vectorial current with the visible sector.
The direct-detection rate is dominated by the interaction with light quarks in this scenario, and we consider the three $\mathrm{SU}(2)\times \mathrm{U}(1)$ invariant operators that couple the first generation of quarks at dimension 6.
Varying the relative magnitude of the three respective Wilson coefficients, we identify regions in the parameter space\,---\,as shown in Fig.~\ref{fig:contours}\,---\,where the direct-detection rate for a xenon target is significantly suppressed.

In models where these Wilson coefficients are generated via $Z^\prime$ exchange, we find that collider and direct-detection experiments have comparable sensitivity for typical choices of the couplings.
For parameter points that are chosen to be comparable to current experimental benchmark scenarios, we find that future direct-detection experiments will test large parts of the parameter space, see e.g.\ Fig.~\ref{fig:axial_Wilson}, that are not accessible at the LHC.

Yet, in a UV-consistent $Z^\prime$ model, anomaly conditions further constrain the allowed parameter space of the Wilson coefficients, and the direct-detection rate can only be suppressed by considering generation-dependent charges.
By charging the right-handed strange quark instead of the right-handed down quark, we find an anomaly-free charge assignment that suppresses the direct-detection rate for xenon targets.
If realised in Nature, such a scenario would have to be tested via collider searches, which could close the window of smaller mediator masses and smaller gauge couplings.

\section*{Acknowledgements}

We would like to thank Joachim Brod, Uli Haisch, Ed Hardy, and Jure Zupan for helpful discussions.
The work of JF and MG has been supported by STFC under the Consolidated Grant ST/T000988/1. This work was also supported by the Deutsche Forschungsgemeinschaft (DFG, German Research Foundation) under Germany's Excellence Strategy - EXC  2121 ``Quantum Universe'' - 390833306 and the Aspen Center for Physics, which is supported by National Science Foundation grant PHY-1607611.

\appendix

\section{Anomaly conditions}
\label{app:anomalies}

We construct here the anomaly-free conditions with only one additional Weyl fermion, and first-generation SM quarks plus potentially one quark flavour from the second-generation. In addition, one generation of leptons must also carry a U$(1)'$ charge in this setup; choosing the third generation helps to evade to some extent the stringent constraints from dilepton searches. The generic U$(1)'$ charge assignment of the right-handed matter fields is given by
\vspace{0.2cm}
\begingroup
\setlength{\tabcolsep}{12pt}
\begin{equation}
\begin{tabular}{c c c c c c c c c} \toprule[1pt]
$Q_{L,1}^c$ & $u_{R,1}$ & $d_{R,1}$  & $L_{L,3}^c$ & $e_{R,3}$ & $\chi_R$ & $u_{R,2}$ or $d_{R,2}$\\\midrule[0.5pt]
$a$ & $b$ & $c$ & $d$ & $e$ & $x$ & $z$\\
\bottomrule[1pt]
\end{tabular}\label{eq:u1p-charges-1g-plus}
\end{equation}
\endgroup
\vspace{0.2cm}
and leads to the following anomaly equations (see Sec. 22.4 of Ref.~\cite{Weinberg:1996kr}):
\begingroup\renewcommand{\arraystretch}{1.3}
\begin{equation}
	\begin{array}{c l}
	\text{SU}(3)\times \text{SU}(3)\times \text{U}(1)':\qquad & 2 a+b+c+z=0 \\
	\text{Grav}\times\text{Grav}\times \text{U}(1)':\qquad &6 a+3 b+3 c+2 d+e+x+3 z=0 \\
	\text{SU}(2)\times \text{SU}(2)\times \text{U}(1)':\qquad &3 a+d=0 \\
	\text{U}(1)_Y\times \text{U}(1)_Y\times \text{U}(1)':\qquad &6 a+48 b+12 c+18 d+36 e+108\,Q_{u(d)}^2\, z=0 \\
	\text{U}(1)_Y\times \text{U}(1)'\times \text{U}(1)':\qquad &-6 a^2+12 b^2-6 c^2+6 d^2-6 e^2+18\,Q_{u(d)} \,z^2=0 \\
	\text{U}(1)'\times \text{U}(1)'\times \text{U}(1)':\qquad &6 a^3+3 b^3+3 c^3+2 d^3+e^3+x^3+3 z^3=0 \\
	\end{array}
\end{equation}
\endgroup

\begingroup
\renewcommand{\arraystretch}{1.5}
\setlength{\tabcolsep}{7pt}
\begin{table}[b]
	\centering
	\begin{tabular}{l c c c c c c c c}\toprule[1pt]
		~ & $Q_{L,1}^c$ & $u_{R,1}$ & $d_{R,1}$ & $u_{R,2}$ & $d_{R,2}$ & $L_{L}^c$ & $e_{R}$ &$\chi_R$\\\midrule[0.5pt]
		S1 &$\frac16(e+x)$&$\frac13(x-2e)$& $\frac13(e-2x)$& $0$ & $0$ & $-\frac12(e+x)$& $e$ &$x$ \\
		S2 &$\frac{1}{4}(x-z)$ & $0$ &$-\frac12(x+z)$ & $z\neq0$ & $0$ &$-\frac34(x-z)$ & $\frac12(x-3z)$ &$x$ \\
		S3 &$\frac12(x+z)$ &$-x-2z$ & $0$ & $0$ & $z\neq0$ & $-\frac32(x+z)$ & $2x+3z$ &$x$ \\
		\bottomrule[1pt]
	\end{tabular}
	\caption{Anomaly-free models where only first-generation quarks or one full generation plus either a right-handed second-generation down or up quarks carry U$(1)^\prime$ carges. Only the left-handed third-generation lepton doublet carries a U$(1)^\prime$ charge to avoid the most stringent bounds from dilepton searches.}
	\label{tab:1g+sr_cr_model}
\end{table}
\endgroup

Here, $Q_{u(d)}$ is the electric charge, $+2/3(-1/3)$, of the quark to which the $U(1)'$ charge, $z$, is assigned in Eq.~\eqref{eq:u1p-charges-1g-plus}. There is one solution with only one-generation, S1, and one solution with an additional second generation up(down)-type quark carrying an independent charge, S2 (S3). These solutions are shown in Table~\ref{tab:1g+sr_cr_model}.  We denote the charge of the new Weyl fermion, $\chi_R$, by $x$, the charge of the right-handed lepton in S1 by $e$, and the charge of the second generation quarks by $z$ in S2 and S3.

\section{A possible UV completion}
\label{sec:uv-model}

We want a DM candidate with purely axial-vectorial coupling to a spin-1 mediator. Since the most minimal additional matter field content\,---\,one Weyl fermion charged under a spontaneously broken $\textrm{U}(1)^\prime$ gauge group\,---\, gives rise to a Majorana fermion after spontaneous symmetry breaking, the desired axial-vectorial coupling is automatically guaranteed.
Hence, we extend the SM gauge group by an additional $\textrm{U}(1)^\prime$ which is spontaneously broken by the vacuum expectation value of a scalar field, $S$, and add one Weyl fermion, $\chi_R$, that is charged under this $\textrm{U}(1)^\prime$ and is additionally odd under a $\mathbb{Z}_2$ symmetry that remains exact. This fermion is neutral under the SM gauge group.

Since the DM candidate $\chi_R$ is chiral under the $\textrm{U}(1)^\prime$, the gauge symmetry is anomalous. One simple solution to make it anomaly free is to also charge one generation of right-handed SM fermions under the $\textrm{U}(1)^\prime$:
\begingroup
\setlength{\tabcolsep}{12pt}
\begin{equation}
	\begin{tabular}{c c c c}\toprule[1pt]
		$u_R$ & $d_R$ & $e_R$ & $\chi_R$\\\midrule[0.5pt]
		$-1$ & $+1$ & $+1$ & $-1$\\
		\bottomrule[1pt]
\end{tabular}
\label{eq:charges}
\end{equation}
\endgroup
\vspace{0.1cm}

\noindent This assignment is sufficient to cancel all mixed and pure anomalies, and it corresponds to the solution S1 in Appendix~\ref{app:anomalies} with $e=1, x=-1$.
However, charging the right-handed SM fermions under $\textrm{U}(1)^\prime$ forbids their Yukawa terms at dimension four. To write these terms, one needs to include powers of the $\textrm{U}(1)^\prime$ Higgs, $S$, suppressed by the same power of the scale $M_*$ where the interaction is generated.

\begin{figure}\centering
	\begin{tikzpicture}[line width=1.2 pt,radius=2pt]
	\def\ri{1};\def\rj{2};\def\dx{0.1};
	\node[label={[label distance=-1mm]120:$u_L^\dagger$}] (tl) at (0,{+\ri}){};
	\node[label={[label distance=-1mm]60:$u_R$}] (tr) at ({6*\ri/2},{+\ri}){};
	\node[label={[label distance=-1mm]240:$\widetilde{H}$}] (bl) at (0,{-\ri}){};
	\node[label={[label distance=-1mm]-60:$S$}] (br) at ({6*\ri/2},{-\ri}){};
	\node (vl) at ({\ri/2},0)  {};
	\node (vm) at ({3*\ri/2},0){};
	\node (vr) at ({5*\ri/2},0){};
	\draw ($(vm)-(0.1,+0.1)$)-- ($(vm)+(0.1,+0.1)$);
	\draw ($(vm)-(0.1,-0.1)$)-- ($(vm)+(0.1,-0.1)$);
	\draw [dashed] (bl.center) -- (vl.center);
	\draw [dashed] (br.center) -- (vr.center);
	\draw[fermion] (vl.center) -- (tl.center);
	\draw[fermion] (vr.center) -- (tr.center);
	\draw[fermion] (vl.center) -- (vm.center);
	\draw[fermion] (vr.center) -- (vm.center);
	\node[label=below:$U^\dagger_L$] at ($(vl)!0.5!(vm)$) {};
	\node[label=below:$U^{\phantom{\dagger}}_R$] at ($(vm)!0.5!(vr)$) {};
	\filldraw (vl) circle;\filldraw (vr) circle;
	\end{tikzpicture}
	\begin{tikzpicture}[line width=1.2 pt,radius=2pt]
	\def\ri{1};\def\rj{2};\def\dx{0.1};
	\node[label={[label distance=-1mm]120:$d_L^\dagger$}] (tl) at (0,{+\ri}){};
	\node[label={[label distance=-1mm]60:$d_R$}] (tr) at ({6*\ri/2},{+\ri}){};
	\node[label={[label distance=-1mm]240:$H$}] (bl) at (0,{-\ri}){};
	\node[label={[label distance=-1mm]-60:$S^\dagger$}] (br) at ({6*\ri/2},{-\ri}){};
	\node (vl) at ({\ri/2},0)  {};
	\node (vm) at ({3*\ri/2},0){};
	\node (vr) at ({5*\ri/2},0){};
	\draw ($(vm)-(0.1,+0.1)$)-- ($(vm)+(0.1,+0.1)$);
	\draw ($(vm)-(0.1,-0.1)$)-- ($(vm)+(0.1,-0.1)$);
	\draw [dashed] (bl.center) -- (vl.center);
	\draw [scalarnoarrow] (br.center) -- (vr.center);
	\draw[fermion] (vl.center) -- (tl.center);
	\draw[fermion] (vr.center) -- (tr.center);
	\draw[fermion] (vl.center) -- (vm.center);
	\draw[fermion] (vr.center) -- (vm.center);
	\node[label=below:$D^\dagger_L$] at ($(vl)!0.5!(vm)$) {};
	\node[label=below:$D^{\phantom{\dagger}}_R$] at ($(vm)!0.5!(vr)$) {};
	\filldraw (vl) circle;\filldraw (vr) circle;
	\end{tikzpicture}
	\begin{tikzpicture}[line width=1.2 pt,radius=2pt]
	\def\ri{1};\def\rj{2};\def\dx{0.1};
	\node[label={[label distance=-1mm]120:$\chi_R$}] (tl) at (0,{+\ri}){};
	\node[label={[label distance=-1mm]60:$\chi_R$}] (tr) at ({6*\ri/2},{+\ri}){};
	\node[label={[label distance=-1mm]240:$S$}] (bl) at (0,{-\ri}){};
	\node[label={[label distance=-1mm]-60:$S$}] (br) at ({6*\ri/2},{-\ri}){};
	\node (vl) at ({\ri/2},0)  {};
	\node (vm) at ({3*\ri/2},0){};
	\node (vr) at ({5*\ri/2},0){};
	\draw ($(vm)-(0.1,+0.1)$)-- ($(vm)+(0.1,+0.1)$);
	\draw ($(vm)-(0.1,-0.1)$)-- ($(vm)+(0.1,-0.1)$);
	\draw [scalarnoarrow] (bl.center) -- (vl.center);
	\draw [scalarnoarrow] (br.center) -- (vr.center);
	\draw[fermion] (vl.center) -- (tl.center);
	\draw[fermion] (vr.center) -- (tr.center);
	\draw[fermion] (vl.center) -- (vm.center);
	\draw[fermion] (vr.center) -- (vm.center);
	\node[label=below:$X^\dagger_L$] at ($(vl)!0.5!(vm)$) {};
	\node[label=below:$X^{\phantom{\dagger}}_R$] at ($(vm)!0.5!(vr)$) {};
	\filldraw (vl) circle;\filldraw (vr) circle;
	\end{tikzpicture}
	\caption{Generating the dimension-5 Yukawa interactions via vector-like-fermions. The fermion flow reflects the fact that we work with $(0,\tfrac{1}{2})$-representation fermions, see~\cite{Dreiner:2008tw}. The diagram for the electron is omitted but it can be obtained from the down-quark one by the replacement $d\to e$ and $D\to E$. The vector-like fermions, $U,D,E,X$ are neutral under $\textrm{U}(1)^\prime$.}
	\label{fig:vector-like-fermions}
\end{figure}
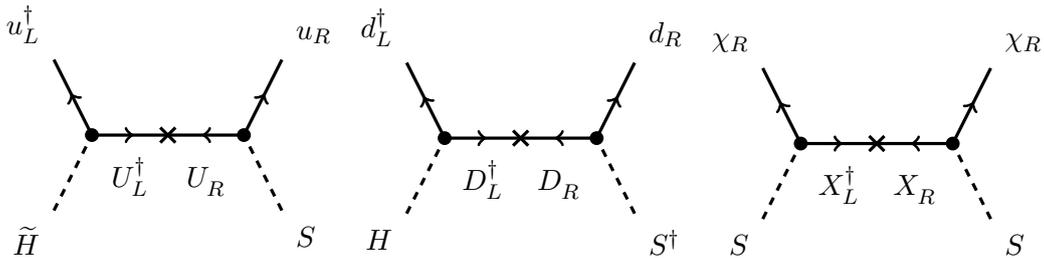

Writing in terms of Weyl spinor fields transforming under the $(0,\tfrac{1}{2})$ representation of the Lorentz group and following the conventions of Ref.~\cite{Dreiner:2008tw}, the Lagrangian describing the fields charged under the $\textrm{U}(1)^\prime$ is given by
\begin{equation}
\mathcal{L}_{\textrm{U}(1)^\prime} = \sum_{f=u,d,e,\chi}\,f_R^\dagger\,iD_\mu\sigma^\mu \,f_R
+\left(D_\mu S\right)^\dagger D^\mu S
-\left[\frac{1}{2} y_\chi\,\chi_R\chi_R\,S + h.c. \right]\,,
\label{eq:u1-lagrangian}
\end{equation}
where $D_\mu = \partial_\mu+i\,g\,q^\prime\,Z^\prime_\mu$ is the $\textrm{U}(1)^\prime$ covariant derivative and $\sigma_\mu\equiv (\mathbb{1}_{2\times 2};\vec{\sigma})$ where $\sigma_i\;\forall\;i\in\{1,2,3\}$ are the Pauli matrices, see Ref.~\cite{Dreiner:2008tw} and references therein.
In order to be able to write the Yukawa term in the square bracket, the $\textrm{U}(1)^\prime$ charge of the Higgs field $S$ must be $+2$. However, such a choice would require one additional $\textrm{U}(1)^\prime$ charged scalar with charge $+1$ to allow for the SM Yukawa terms given the charge assignment~\eqref{eq:charges}.
Thus, a solution that would allow all Yukakwa interactions with only one $\textrm{U}(1)^\prime$-charged scalar, $S$, forces us to assign it a charge $+1$. This choice forbids the Yukawa term in Eq.~\eqref{eq:u1-lagrangian} at the renormalizable level and all Yukawa terms now arise at dimension five in the following way (again, writing in terms of fields that transform under the $(0,\tfrac{1}{2})$ representation of the Lorentz group as before)

\begin{equation}
\mathcal{L}_{\text{U}(1)^\prime}^{\text{Yukawa}} =
- y_u\,\widetilde{H}\cdot Q_L^\dagger u_R\,\frac{S}{M_*}
- y_d\,H\cdot Q_L^\dagger d_R\,\frac{S^\dagger}{M_*}
- y_e\,H\cdot L_L^\dagger e_R\,\frac{S^\dagger}{M_*}
- y_\chi \,\chi_R\chi_R\,\frac{S^2}{M_*^2}
+ \text{h.c.}\,,
\label{eq:yukawas}
\end{equation}
where $\widetilde{H}^a=\epsilon^{ab} H^{\dagger}_b$ and $a,b$ are $\textrm{SU}(2)_L$ indices which are made explicit here for clarity while they are suppressed in the equation above where their contraction via the $\delta_a^b$ invariant tensor is denoted by $X\cdot Y\equiv \delta_a^b\,X^a Y_b$ with $X$ and $Y$ transforming under (formally) conjugate representations.
These higher dimensional operators can be generated at the scale $M_*$ via vector-like fermions with masses of $\mathcal{O}(M_*)$ as shown in Fig.~\ref{fig:vector-like-fermions}.
For each of the fermions of~\eqref{eq:charges}, we require one pair of vector-like Weyl fermions which are neutral under $\textrm{U}(1)^\prime$ but are otherwise charged under $\textrm{SU}(3)_C$ or $\textrm{U}(1)_Y$ as necessary. The vector-like fermion $X$ corresponding to the DM candidate $\chi_R$ is completely neutral under the SM and the $\textrm{U}(1)^\prime$ gauge group. However, it must also be odd under $\mathbb{Z}_2$ in order for the DM Yukawa term to respect it.

After spontaneous symmetry breaking of the $\rm{U}(1)^\prime$, the DM candidate $\chi_R$ transforms only under the $\mathbb{Z}_2$ symmetry as $\chi_R\to -\chi_R$ but does not carry any additional conserved charges. Thus, we can construct a left-handed, $(\tfrac{1}{2},0)$, fermion $\epsilon_{\alpha\beta}(\chi_R^\dagger)^{\beta}$ with the same quantum numbers as $\chi_{R}$ and, consequently, we can construct a four-component Majorana spinor as
\begin{eqnarray}
\chi_M = \begin{pmatrix}
\big[\chi_{R}^{\dagger}\big]_\alpha\\
\big[\chi_{R}\big]^{\dot{\alpha}}
\end{pmatrix}\,,
\end{eqnarray}
which explicitly satisfies the Majorana ``reality'' condition $\chi_M^c = \chi_M$, though it is manifestly obvious that this must be so since the four-component spinor is constructed from only one Weyl fermion.
The Lagrangian of this Majorana DM is given by
\begin{equation}
\lag_M = \frac{1}{2}\bar{\chi}_M\,i\slashed{\partial}\,\chi_M
+\frac{1}{2}\bar{\chi}_M\,\gamma^\mu\gamma_5\,\chi_M\,Z^\prime_\mu
-\frac{1}{2}m\,\bar{\chi}_M\,\chi_M\,.
\end{equation}

%%%%%%%%%%%%%%%%%%%%%

\bibliographystyle{JHEP}
\bibliography{DD_vs_collider}

%%%%%%%%%%%%%%%%%%%%%

\end{document}